\documentclass[aps,pra,preprint,showpacs,preprintnumbers,amsmath,amssymb,footinbib]{revtex4}
\usepackage{graphicx}
\usepackage{bm}
\usepackage{hyperref}
\usepackage[final]{changes}
\begin{document}
\preprint{}
\title{The Eccentric Collective BFKL Pomeron}

\author{Larry McLerran } 

\affiliation{RIKEN BNL, Brookhaven National Laboratory, 
Upton, NY 11973 \\ 
Physics Department, Brookhaven National Laboratory,
Upton, NY 11973, USA \\
Physics Department, China Central Normal University, Wuhan, China}

\author{Vladimir V. Skokov}
\email{vladimir.skokov@wmich.edu}
\affiliation{Department of Physics, Western Michigan University, Kalamazoo, MI 49008, USA}

\begin{abstract}
We apply the flow analysis for multi-particle correlations used in heavy ion collisions
to multi-particle production from a  Pomeron.  We show that the $n$'th order angular harmonic
arising from an $m$ particle correlation $v_n[m]$  satisfies $v_n[m] \approx v_n[p]$ for $n \ge 1$.  We discuss
some implications of this for the Color Glass Condensate description of high energy hadronic collisions.
\end{abstract}

\maketitle

\section{Introduction}

\begin{figure}
\includegraphics[width=0.3\linewidth]{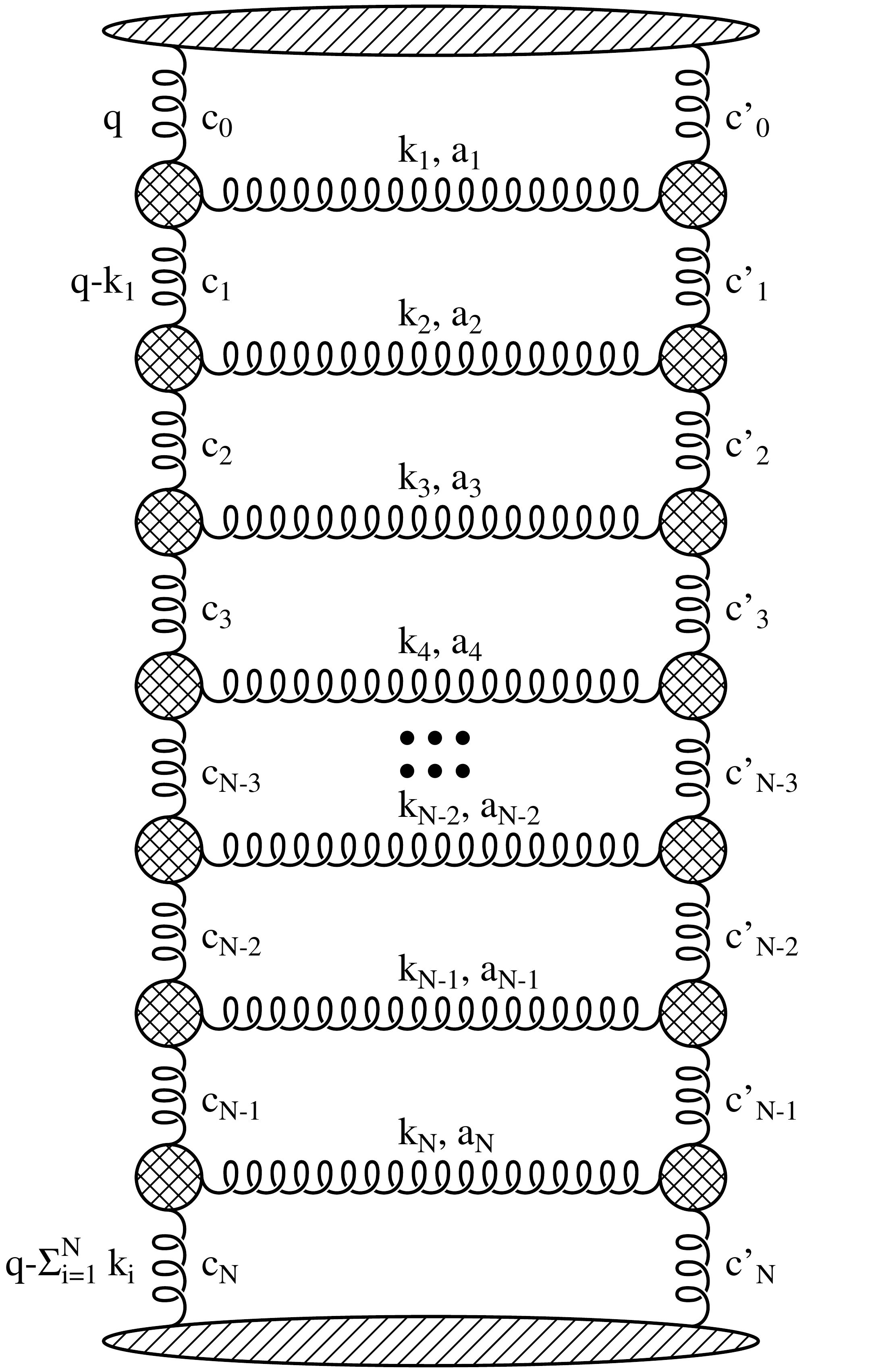}
\caption{$N$-gluon production. The shaded blob denotes Lipatov vertex.  }
\label{fig:Ngluon}
\end{figure}

The BFKL pomeron is presumably responsible for driving the high energy growth of cross sections
in high energy hadronic collisions~\cite{Balitsky:1978ic}.  In parton-parton scattering, the Pomeron 
would correspond to the ladder graph diragram of Fig.~\ref{fig:Ngluon}.  In Fig.~\ref{fig:Ngluon}, we will consider
the imaginary part of this diagram corresponding to multi-gluon production.  For such a process the momentum 
of the particles initiating the Pomeron exchange  at the top and the bottom of the diagram are equal in the initial and final
state, so that the momentum on the struts of the ladder are equal.  
If the momentum transfer imparted to the struts, $q$, is large, this diagram can be evaluated in weak coupling and gives the perturbative
BFKL pomeron.  The BFKL pomeron leads to evolution of quark and gluons distribution functions through the BFKL
equation~\cite{Balitsky:1978ic}.

In theories of gluon saturation~\cite{Gribov:1984tu,Mueller:1985wy,McLerran:1993ni,McLerran:1993ka}, the momentum 
scale associated with Pomeron exchange is that of the saturation momenta, and the evolution equation 
for the saturation momentum is basically derivative of that of evolution of the BFKL pomeron. 
The basic content of the Color Glass
Condensate description of such processes is that the sources at the top and bottom of the ladder 
are replaced by a distribution of colored sources.  These color sources are coherent, so that the 
infrared integrations over momentum transfer are cut-off at the saturation momenta of the upper and
the lower hadron participating in the collision.  When computing multi-particle production one is
determining inclusive particle production with such a diagram, and one should look only over a finite 
range of rapidity between the upper produced gluon and the lower produced gluon.  The saturation momentum 
of the upper hadron is at that of the upper rapidity and similarly for the lower hadron.  The restriction 
on the total rapidity is that it is of order $\alpha_s N_c \Delta y \ll 1$

\begin{figure}
\includegraphics[width=0.5\linewidth]{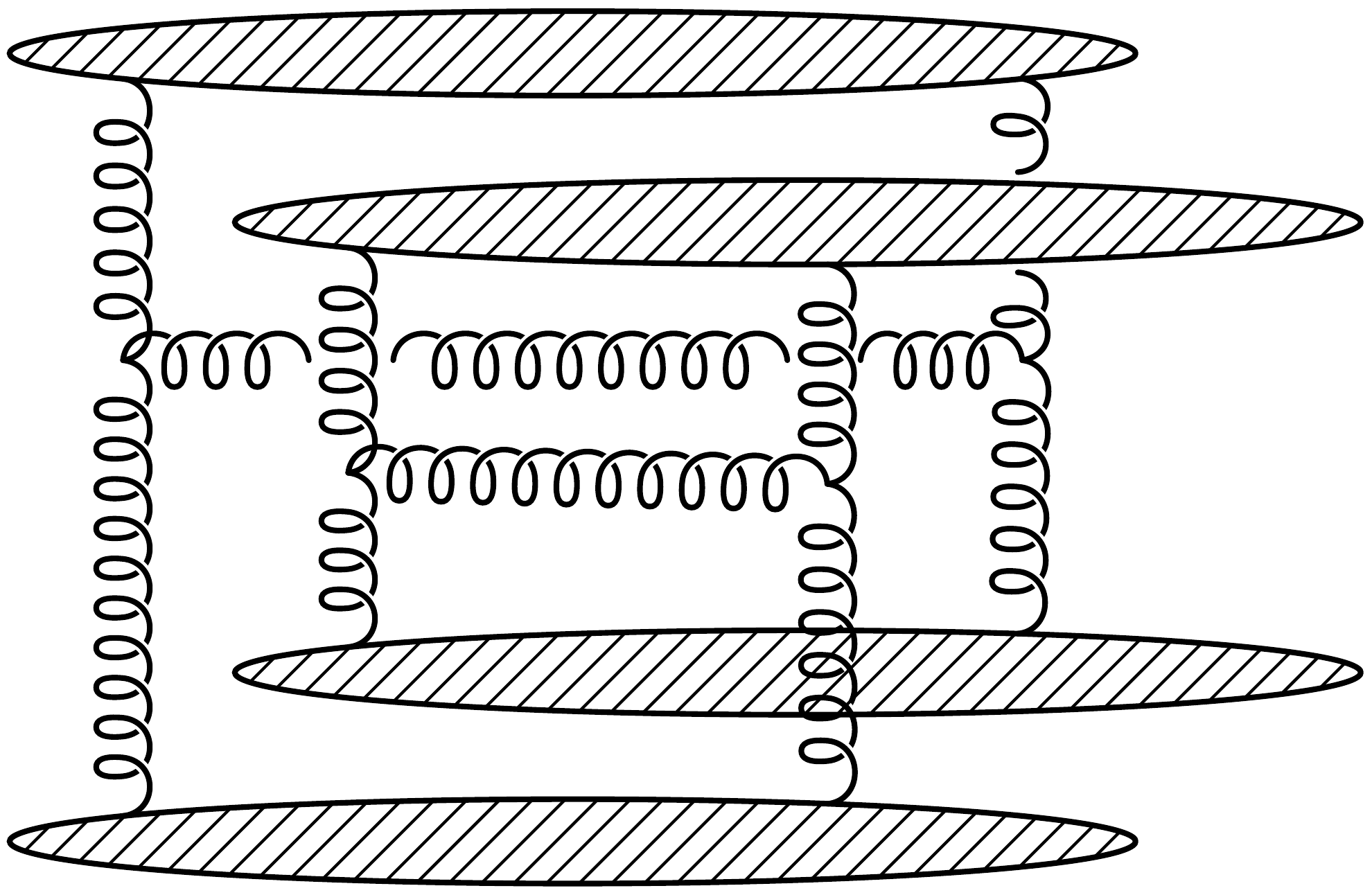}
\caption{The glasma graph.  }
\label{fig:glasma}
\end{figure}

When one considers multi-gluon correlations, there are a variety of possible effects. 
In this paper we will compute the contribution arising from the Pomeron.  
In general in hadron collisions, there are contributions from final state interactions.
In heavy ion collisions, AA~\cite{Alver:2010gr,Adams:2004bi} and also perhaps high
multiplicity pp~\cite{Khachatryan:2010gv} and pA~\cite{CMS:2012qk} these may be the dominant 
effects~\cite{Bozek:2012gr}.  In addition there are two particle correlations generated by 
the initial state of such collisions.  In particular there is the two particle correlation generated in the CGC. 
The diagram of Fig. \ref{fig:glasma} 
generates such a correlation~\cite{Dumitru:2008wn,Dumitru:2010iy,Dusling:2012iga}. 
It is of leading order in the classical approximation but suppressed by $1/N_c^2$ in the large number of colors limit.

The diagrams we consider arising from Pomeron exchange are of higher order in $\alpha_s$ than are
the two particle correlation diagrams usually considered for the Color Glass Condensate.  
They are however a factor of $N_c^2$ larger.  Note also that in the case pp or pA collisions,
the Pomeron diagram is also enhanced due to density factors associated with the coupling of the
correlated diagram to the CGC.  For example, if one couples to a dilute projectile with a 
classical field strength of order $g$, and a dense particle target with strength $1/g$, we have the following cases
for two gluon production
\begin{center}
\begin{tabular}{ c | c | c }
 System &  Pomeron graph &  Glasma graph \\
 \hline
 Dense-Dense & 1 &  $1/g^4$ \\
 Dilute-Dense & $g^4$ & $g^4$\\
 Dilute-Dilute & $g^8$ & $g^{12}$ 
\end{tabular}
\end{center}

Such counting is of course simplistic, since if we are at small enough momenta, we are in the region
where the saturation momenta of the proton is important, and then the counting of powers of $g$ 
in proton-proton collisions is similar to that of heavy ions.  Another subtlety for heavy-ion collisions
or high multiplicity events is that the overall normalization of the flow
contributions is scaled by the  production cross section with no angular dependence, and this contains 
contributions from multiple particle processes that involve many gluon exchanges, and are in addition to the
contribution of the Pomeron.
To properly compute the factors associated with the typical Glasma diagram and that of the Pomeron is of 
course not so easy to do, but our point is that we might expect the Pomeron to play an increasingly 
important role in $pp$ and $pA$ collisions relative to $AA$. 

In any case, how one resolves the Pomeron and separates it from other effects is not the goal of this paper.  
Our much more modest goal is to explore the multiparticle correlations associated with the Pomeron decay into gluons.
We find that there is a rich structure of correlations, and in particular, we find that the multiparticle moments
satisfy $v_n[p] \approx v_n[m]$.  This approximate equality is a signature of the collectivity of motion of 
the gluons produced from a single Pomeron.  It's origin is not hydrodynamical but is associated with the coherence
of the underlying emission process.  It might be possible to isolate and study such processes in electron-positron
annihilation experiments and in deep inelastic scattering experiments.

\section{Angular Correlations in Gluon Bremsstrahlung}

We first analyze the two particle correlation induced in a single Pomeron decay.  Our analysis
parallels the insightful work Gyulassy, Levai, Vitev and Csorgo~\cite{Gyulassy:2014cfa} 
of the underlying process of gluon bremsstrahlung first analyzed by Bertsch and Gunion~\cite{Gunion:1981qs}.

The formula for multiple gluon production using the Lipatov vertex formalism valid in large $N_c$ 
for high energy multi gluon production.

\begin{eqnarray}
\frac{d^{3N} \sigma } {dy_1 d^2 k_1 dy_2 d^2 k_2  \cdots dy_N d^2 k_N } && = f 
\int{d^2 q_\perp}
\frac{1}{ {q}_\perp^2 +\mu^2 }  \frac{1}{\left( \vec{q}_\perp - \sum_{j=1}^n \vec{k}_{\perp j} \right)^2+\mu^2} \prod_{i=1}^N \frac{1}{k_{\perp i}^2}  
\label{Ngluon}
\end{eqnarray}
The overall factor $f$ depends on the particular system, for dilute-dilute scattering 
$f= \frac {1}{2} (4 g^2)^{N+2}  C_A^N C_F N_c$, where $C_A=N_c$ and $C_F = \frac{N_c^2-1}{2N_c}$. 

In this formula $N$ particles are produced with transverse momenta $k_i$ and rapidity $y_i$.  
The factors of $\mu^2$ in this cross section are infrared cutoff squared, which in the saturation picture,
is  the saturation momentum, $Q_s$

A useful relationship that later will be  applied  in our analysis is
\begin{equation}
1 = \int d^2 p_\perp \delta( \vec{p}_\perp + \vec{q}_\perp - \sum_i \vec{k}_{\perp i})  = \int \frac{d^2 p_\perp}{(2\pi)^2} \int d^2 x_\perp 
e^{i\vec{x}_\perp(\vec{p}_\perp + \vec{q}_\perp - \sum_i \vec{k}_{\perp i})}
\label{Eq:Fourier}
\end{equation}
Use of this relationship allows integration over the final-state momenta
in a way that exploits the fundamental factorization of the integrated 
$N$ particle production amplitude.

\section{2 particle production}

Let us begin by computing $v_n$ for the two particle amplitude.  We first write down a formula for the integrated
two particle correlation projected onto an angular dependence $e^{in\phi}$
\begin{eqnarray}
&&\frac{d^2 \sigma_n }{d y_1 d y_2} = 
f 
\int \frac{d^2 k_{1 \perp }}{k_{1 \perp }^2} e^{i\phi_1 n} \int   \frac{d^2 k_{2 \perp }}{k_{2 \perp }^2}  e^{-i\phi_2 n} 
\int \frac{d^2 q_\perp}{q_\perp^2+\mu^2} \frac{1}{(\vec{q}_\perp - \vec{k}_{1 \perp } - \vec{k}_{1 \perp })^2+\mu^2} = 
\nonumber 
\\ 
&& = f  \int \frac{d^2 x_\perp}{(2\pi)^2} \left( \int \frac{d^2 q_\perp}{q_\perp^2+\mu^2}e^{i \vec{x}_\perp \vec{q}_\perp} \right)^2 \int \frac{d k_{1 \perp }}{k_{1 \perp }} \int \frac{d k_{2 \perp }}{k_{2 \perp }} \times
\nonumber
\\ 
&&\times \int d \phi_1 e^{i(\phi_1 n - x_\perp k_{1\perp} \cos( \phi_1 ))}   e^{i \phi_x n}   \int d \phi_2 e^{i(-\phi_2 n - x_\perp k_{2\perp} \cos( \phi_2 ))}  e^{-i \phi_x n} =
\nonumber
\\ 
&& = f \int \frac{d^2 x_\perp}{(2\pi)^2} \left( 2\pi K_0(\mu x_\perp) \right)^2 
\int \frac{d k_{1 \perp }}{k_{1 \perp }} \int \frac{d k_{2 \perp }}{k_{2 \perp }}  \left[  2\pi (-i)^n J_n(x_\perp k_{1 \perp }) 
\right] \left[ 2\pi (-i)^n J_n(x_\perp k_{1 \perp }) )\right]         = \nonumber \\
&&   = \frac{(2\pi)^3 f }{2 \mu^2} \frac{(-1)^n}{n^2}
\label{v2n}
\end{eqnarray}
To derive this we first applied Eq.~\eqref{Eq:Fourier} and then used the following well known integrals
\begin{eqnarray}
&&\int \frac{d^2 q_\perp}{q_\perp^2+\mu^2}e^{i \vec{x}_\perp \vec{q}_\perp}  = 2\pi K_0(\mu x_\perp), \\ 
&&\int d \phi_1 e^{i(\phi_1 n - x_\perp k_{1\perp} \cos( \phi_1 ))} =  2\pi (-i)^n J_n(x_\perp k_{1 \perp }).  
\label{Bessels}
\end{eqnarray}
where $K$ and $J$ are corresponding Bessel functions. For an integer $n$, $J_n(x)=(-1)^n J_{-n}(x)$.
The above expression Eq.~\eqref{v2n} is only true for $n>0$. For $n=0$, the integral
$ \int \frac{d k_{ \perp }}{k_{ \perp }}  J_0(x_\perp k_{ \perp }) $ is divergent and should be 
properly regularized at the saturation momentum $\mu$:
\begin{eqnarray}
\int_{\mu}^\infty \frac{d k_{ \perp }}{k_{ \perp }}  J_0(x_\perp k_{ \perp }) &=&  
\lim_{\varepsilon \to 0}
\left(
\int_0^\infty k^\varepsilon \frac{d k_{ \perp }}{k_{ \perp }}  J_0(x_\perp k_{ \perp })  
- \int_0^\mu k^\varepsilon \frac{d k_{ \perp }}{k_{ \perp }}  J_0(x_\perp k_{ \perp })  
\right) = \\ &&= \ln 2 - \gamma_E - \ln(\mu x_\perp). 
\label{RegulJ0}
\end{eqnarray}
Using this result we obtain for $n=0$:
\begin{eqnarray}
&&\frac{d^2 \sigma_0 }{d y_1 d y_2} = 
f 
\int \frac{d^2 k_{1 \perp }}{k_{1 \perp }^2}  \int   \frac{d^2 k_{2 \perp }}{k_{2 \perp }^2}  
\int \frac{d^2 q_\perp}{q_\perp^2+\mu^2} \frac{1}{(\vec{q}_\perp - \vec{k}_{1 \perp } - \vec{k}_{1 \perp })^2+\mu^2} = 
\nonumber 
\\ 
&& = f \int d^2 x_\perp \left( 2\pi K_0(\mu x_\perp) \right)^2 
\left(
\ln 2 -\gamma_E - \ln(\mu x_\perp)
\right)^2
= \nonumber \\
&&   = (2\pi)^3 f \frac{1}{\mu^2}
\label{v20}
\end{eqnarray}
Here and later we will use the following integral
\begin{equation}
S_m = \int dx x (K_0(x))^2 (\ln(2)-\gamma_E - \ln(x))^m.  
\end{equation}
The analytic expression for $S(m)$ is derived in Appendix 1. Here we note that  $S_0=1/2$ and  $S_2=1$.
Obviously in this article, we use $S_m$ only for even $m$.

Thus for $\langle v_n^2 \rangle $ we have 
\begin{equation}
\langle v_n^2 \rangle = \frac{\frac{d^2 \sigma_n }{d y_1 d y_2}} {  \frac{d^2 \sigma_0 }{d y_1 d y_2}   } = \frac{(-1)^n}{2 n^2}.
\end{equation}
The flow cumulants for 2-particle correlation are defined according $v_n[2] = \sqrt{\langle v_n^2 \rangle    }$.
Thus, e.g. for $n=2$,  $v_n[2] = \frac{\sqrt{2}}{4}$.   The factor of $-1$ for the odd $<v_n^2>$
is a consequence of the backard peaking of the two particle correlation, and is ditinctively different
from hydrodynamical flow indeuced correlations.


The two particle correlation function $C_2(\Delta \phi)$ 
defined by 
\begin{equation}
C_2(\Delta \phi) = \left(   \frac{d^2 \sigma_0 }{d y_1 d y_2}  \right)^{-1} 
\int \frac{d^2 k_{1 \perp }}{k_{1 \perp }^2}  \int   \frac{d^2 k_{2 \perp }}{k_{2 \perp }^2}  
\delta(\Delta\phi +\phi_1-\phi_2)  \frac{d^6 \sigma }{d y_1 d^2 k_{\perp 1} d y_2  d^2 k_{\perp 2}  }. 
\label{C2}
\end{equation}
can be analytically computed using
\begin{equation}
\delta(x) = \frac{1}{2\pi}\sum_{a=-\infty}^\infty e^{i a x}. 
\end{equation}

Applying the same transformations as for $d^2 \sigma_n/dy_1 dy_2$ we arrive to 
\begin{eqnarray}
C(\Delta \phi) = 1 + \frac{1}{2\pi} \sum_{a=1}^\infty \frac{(-1)^a}{a^2} \cos(a\Delta \phi ) = 1 -\frac{\pi}{24} + \frac{1}{8\pi} \Delta \phi^2.
\label{CorrFun}
\end{eqnarray}
This equation is correct for $|\Delta \phi| < \pi$.

\begin{figure}
\includegraphics[width=0.5\linewidth]{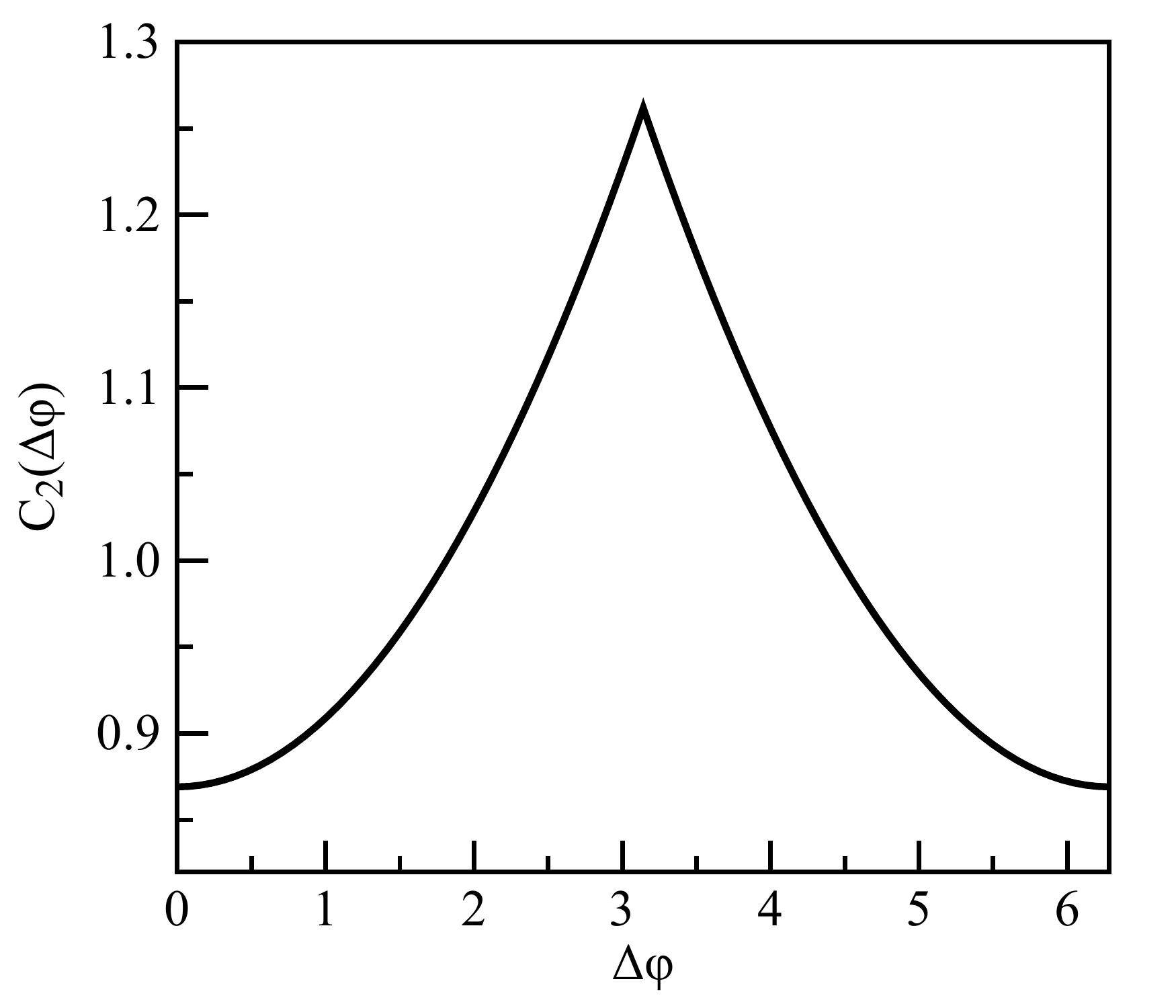}
\caption{$C_2(\Delta \phi)$ for two particle production.}
\label{fig:BFKLCorr}
\end{figure}

\section{$2m$ particle production}
Analogously to Eq.~\eqref{v2n}
\begin{eqnarray}
\frac{d^{2m} \sigma_n  } {d  y_{ 1} d  y_{ 2} \cdots  d  y_{ 2m}  } 
&=& f  \int \frac{d^2 k_{1 \perp }}{k_{1 \perp }^2} e^{i\phi_1 n} 
\int  \frac{d^2 k_{2 \perp }}{k_{2 \perp }^2}  e^{i\phi_2 n} 
\cdots 
\int \frac{d^2 k_{m \perp }}{k_{m \perp }^2} 
e^{i\phi_m n} \times
\\ && \int \frac{d^2 k_{m+1 \perp}}{k_{m+1 \perp}^2}  e^{-i\phi_{m+1} n} 
\cdots \int \frac{d^2 k_{2m \perp }}{k_{2m \perp}^2}  e^{-i\phi_{2m} n} \times
\\ &&\int \frac{d^2 q_\perp}{q_\perp^2+\mu^2} \frac{1}{(\vec{q}_\perp - \sum_{i=1}^m \vec{k}_{i \perp})^2+\mu^2} =\\
&& (-1)^{n\cdot  m} \left(\frac{2\pi}{n} \right)^{2m}  \frac{2\pi f  }{2 \mu^2}. 
\label{Eq:v2m}
\end{eqnarray}
for non-zero $n$. For $n=0$ we get  
\begin{eqnarray}
\frac{d^{2m} \sigma_0  } {d  y_{ 1} d  y_{ 2} \cdots  d  y_{ 2m}  } 
&=& f  \int \frac{d^2 k_{1 \perp }}{k_{1 \perp }^2} 
\int  \frac{d^2 k_{2 \perp }}{k_{2 \perp }^2}  
\cdots 
\int \frac{d^2 k_{m \perp }}{k_{m \perp }^2} 
 \times
\\ && \int \frac{d^2 k_{m+1 \perp}}{k_{m+1 \perp}^2}  
\cdots \int \frac{d^2 k_{2m \perp }}{k_{2m \perp}^2}  \times
\\ &&\int \frac{d^2 q_\perp}{q_\perp^2+\mu^2} \frac{1}{(\vec{q}_\perp - \sum_{i=1}^m \vec{k}_{i \perp})^2+\mu^2} =\\
&&  \left( 2\pi \right)^{2m+1}  \frac{f S_{2m} }{\mu^2}. 
\label{Eq:v2nm}
\end{eqnarray}
Thus 
\begin{equation}
\langle v_n^m \rangle =  \left(  \frac{d^{2m} \sigma_0  } {d  y_{ 1} d  y_{ 2} \cdots  d  y_{ 2m}  } 
 \right)^{-1}  \frac{d^{2m} \sigma_n  } {d  y_{ 1} d  y_{ 2} \cdots  d  y_{ 2m}  } 
 = \frac{1}{2 S_{2m}} \frac{(-1)^{nm}}{n^{2m}}. 
\label{vnm}
\end{equation}
The corresponding cumulants $v_n[2m]$ can be computed using expressions given in Appendix 2.
Here we provide the numerical values:
\begin{equation}
\begin{array}{ll}
 v_2[2] =& 0.353553, \\
 v_2[4] =& 0.404931, \\
 v_2[6] =& 0.40857, \\
 v_2[8] =& 0.408991, \\
 v_2[10] =& 0.409049, \\
 v_2[12] =& 0.409057.
\end{array}
\end{equation}

\section{Summary and Conclusions}

We have shown that the flow analysis of the Pomeron indicates a pattern of coherence which certainly
indicates collective motion of its decay products.  We have been careful to restrict our consideration to 
processes where single Pomeron exchange is the dominant contribution.  One might ask if it might be relevant for heavy ion collisions.  In the two particle correlations of pA or pp collisions, the jet contribution is explicitly
subtracted.  If this is properly done the, the Pomeron contribution should be removed, and the remainder
is the diagram of Fig. 2.  In addition, in pA or pp  collisions, there are all possible manners of final
state interactions which might generate collective effects.

In multi-particle correlations with numbers of particles greater than 2, no  subtraction of the jet contributions done for pA collisions, so the Pomeron might make a significant contribution.  However, it is important to remember that the collectivity of the Pomeron
is really associated with a backwards recoil peak for the Pomeron as is shown in Fig. 3.  This means that
the computed $(v_n(4p+2))^{4p+2}$ would be negative for odd $n$.  The first place this would appear would be in
in $v_3[6]$.  A measurement of such a correlation would give a solid measure of whether or not the collectivity
in pA collisions arises from coherence of Pomeron decays or other effects.

The coherence seen in the Pomeron decay suggests that there will be entirely non-trivial coherence patterns
in other multi-particle processes.  We would associate such coherence with an initial state effect.  Perhaps
something along the lines of Ref.~\cite{Gyulassy:2014cfa} or of Ref.~\cite{Dumitru:2014dra} are steps in the correct direction for making a theory.

Even if there is little impact of these results for pA or pp collisions, the coherence of the decay products
observed for the Pomeron may have implications for elementary processes such as jet decay in $e^+e^-$
annihilation of in deep inelastic scattering.  A proper determination of such effects would require
an analysis of the fragmentation of the gluons produced in such collisions.

\section{Acknowledgements}

The research of L.M. is supported under DOE Contract No. DE-AC02-98CH10886. 
We thank Francois Gellis for very useful comments. 
Lary McLerran acknowledges very useful 
conversations with Miklos Gyulassy on this subject, and a most stimulating talk he gave at Quark Matter 2014,
where these ideas were initiated.

\section{Appendix 1}

\begin{equation}
S_m = \int  d x  x   K^2_0( x) 
\left(\ln 2 -  \ln(x) - \gamma_E   \right)^{m}
\label{Sm}
\end{equation}
This integral can be taken analytically 
\begin{equation}
S_m = \sum_{i=0}^{m}\binom{m}{i} (\ln 2 -\gamma_E)^{m-i} I_i 
\label{Sm_1}
\end{equation}
where 
\begin{equation}
I_i = \int dx x K^2_0(x) \ln^i(x) = \frac{d^i}{d\alpha^i} \left. \left( \frac{\sqrt{\pi}}{4}  \frac{\Gamma^3\left(\frac{\alpha+1}{2}\right)} {\Gamma(\alpha/2+1)} \right) \right|_{\alpha=1}
\label{I}
\end{equation}
This integral can be derived from 
\begin{equation}
\int K_0^2(x) x^\alpha dx =  \frac{\sqrt{\pi}}{4}  \frac{\Gamma^3\left(\frac{\alpha+1}{2}\right)} {\Gamma(\alpha/2+1)} 
\end{equation}
Some value for $S_m$
\begin{eqnarray}
S_0&=&1/2, \\ 
S_1 &=& 1/2, \\
S_2 &=& 1, \\
S_4 &=& -2 \zeta (3)+12-\frac{\pi ^4}{40}, \\
S_6 &=& -60 \zeta (3)+5 \zeta (3)^2-63 \zeta (5)+360-\frac{3 \pi ^4}{4}-\frac{5 \pi ^6}{84}.
\label{Ss}
\end{eqnarray}
All factors of $\ln 2$ appearing on the right hand side of Eq. \eqref{Sm} cancel 
in the final expressions for $S_m$.  

\section{Appendix 2}
For completeness here we also list  the expressions for the cumulants
\begin{eqnarray}
 v_n[2]^2 &=& \langle v_n^2\rangle , \\ 
 v_n[4]^4 &=&   2\langle v_n^2\rangle ^2-\langle v_n^4 \rangle, \\ 
 v_n[6]^6 &=&  \frac{1}{4} \left(12 \langle v_n^2\rangle ^3-9 \langle v_n^4 \rangle \langle v_n^2 \rangle 
	  + \langle v_n^6 \rangle  \right), \\
 v_n[8]^8 &=& \frac{1}{33} \left(144 \langle v_n^2\rangle ^4-144  \langle v_n^4 \rangle \langle v_n^2 \rangle^2+
			 16  \langle v_n^6 \rangle \langle v_n^2 \rangle 
			 +18  \langle v_n^4 \rangle^2   -  \langle v_n^8 \rangle \right), \\
 v_n[10]^{10} &=& 
				\frac{1}{456} \left(2880  \langle v_n^2\rangle^5-3600 \langle v_n^4 \rangle \langle v_n^2 \rangle 
				  ^3+400 \langle v_n^6 \rangle \langle v_n^2 \rangle ^2+
					        \right.\\&& \left.
									        25 \left(36 \langle v_n^4 \rangle ^2-\langle v_n^8 \rangle \right)\langle v_n^2 \rangle -100 \langle v_n^4 \rangle  \langle v_n^6 \rangle +\langle v_n^{10} \rangle \right)
\end{eqnarray}

\end{document}